\documentclass[12]{iopart}
\usepackage{color,graphicx,times}
\usepackage[applemac]{inputenc}
\usepackage{epsfig}

\begin{document}

\title{ Experimental open air quantum key distribution with a single photon 
source }

 \author {R~All\'eaume$^{\dag}$
, F~Treussart$^{\dag}$, G Messin$^{\ddag}$,  Y Dumeige$^{\dag}$, J-
F~Roch$^{\dag}$, A Beveratos$^{\ddag}$, R Brouri--Tualle$^{\ddag}$, J-
P~Poizat$^{\ddag}$   and P Grangier$^{\ddag}$      }

\address{\dag\ Laboratoire de Photonique Quantique et Moléculaire, UMR 8537 
du CNRS, ENS Cachan,  61 avenue du Président Wilson, 94235 Cachan Cedex 
France}

\address{\ddag\ Laboratoire Charles Fabry de l'Institut d'Optique, UMR 8501 
du CNRS, F- 91403 Orsay France}

\begin{abstract}
We present  a full implementation  of  a quantum key distribution (QKD) system 
with a single photon source, operating at night in open air. The single 
photon source at the heart of the functional and reliable setup relies on 
the pulsed excitation of a single nitrogen-vacancy color center in diamond 
nanocrystal.  We   tested the effect of attenuation on the polarized encoded 
photons  for inferring longer distance performance of our system. For strong 
attenuation, the use of  pure single photon states gives measurable 
advantage over systems relying on weak attenuated laser pulses. The results 
are in good agreement with theoretical models developed to assess QKD 
security.
\end{abstract}

\submitto{New J. Phys.}

\maketitle

\section{Introduction}

Key distribution remains a central problem in cryptography, as encryption 
system security cannot exceed key security. Public key protocols rely on 
computational difficulty \cite{Diffie_76}. They  cannot however guarantee 
unconditional security against future algorithm or hardware advances. 

As Bennett and Brassard first proposed twenty years ago \cite{BB_84}, 
quantum physics can be used to build alternative protocols for key 
distribution \cite{Gisin_RevModPhys}. In their proposed ``BB84''  scheme for quantum key distribution (QKD), a first user (Alice) sends a second user (Bob) a sequence of single photons on an authenticated channel. Each of them is independently and randomly prepared in one of the four polarization states, linear-vertical (V), linear-horizontal (H), circular-left (L), circular-right (R). For each photon he detects, Bob picks randomly one of the two non orthogonal bases to perform a measurement. He keeps the outcome of his measurement secret and Alice and Bob publicly compare their basis choices. They only keep data for which polarization encoding and measurements are done in the same basis. In the absence of experimentally induced errors and eavesdropping, the set of data known by Alice and Bob should agree. Due to quantum physics' contraints on single photon measurements an eavesdropper (commonly named Eve) cannot gain even partial information without disturbing the transmission. The unavoidable errors introduced by Eve can be detected by the legitimate users of the quantum transmission channel. If the measured error rate is too high, no secret can be generated from the transmitted data. But if the error rate remains within acceptable bounds Alice and Bob can distill a secure secret key, perfectly unknown by Eve, using key reconciliation procedures. This perfectly secure key can then be used for data encryption.

 Interest in experimental QKD has evolved from 
early proof-of-principle experiments \cite{Bennett_92,Townsend_94} to long 
distance demonstrations on optical fibers \cite{Ribordy_01,Kosaka_03} as 
well as in free space \cite{Hughes_02,Rarity_02,Kurtsiefer_03}  and now to 
commercial products \cite{magiq_web,idquantique_web}. Nevertheless, several 
technological and theoretical barriers still have to be overcome to improve 
performance of current QKD systems.
Most of them relie on weak coherent pulses (WCP) as an approximation to 
single photons.  Such classical states are very simple to produce  but a 
fraction of them will 
contain two photons or more. Since information exchanges using such 
multiphotonic pulses
can be spied on by potential eavesdropping 
strategies~\cite{Brassard_00,Lutkenhaus_00}, security hazard  is introduced 
in the key distribution process.  For QKD  schemes relying  on  WCP, one has 
finally to throw away  part of the initially exchanged information,   
proportional to what an eavesdropper could have learned from it. Indeed, in 
WCPs' schemes, the probability for multiphotonic pulses is directly 
connected to the mean intensity of the initial pulse that must therefore be 
attenuated more and more to guarantee security as line losses become higher. Therefore either
 transmission rate at long distance becomes vanishingly small 
or complete security cannot be guaranteed.

The use of   true single photon source (SPS)  presents an intrinsic 
advantage over WCPs' schemes since it  potentially permits  greater per-bit 
extraction of secure information. This  advantage becomes significant for 
systems with high losses on the quantum transmission channel like   
envisioned satellite QKD  \cite{Rarity_02}. 
 Single photon quantum cryptography  has recently been implemented in two 
experiments \cite{SPS_alex,SPS_waks} which gave clear evidence for  that  
advantage. Following the work of Beveratos et al~\cite{SPS_alex}, we have 
used  a pulsed SPS, based 
 on   temporal control of the fluorescence of   single color nitrogen 
 vacancy (NV)  center  in   diamond nanocrystal. On the emitted polarized 
photons, we have then    implemented the ``BB84'' QKD protocol \cite{BB_84}. 
 In our realization, quantum communication between Alice and Bob has been 
realized in open air during  night,  between the two wings  of the Institut 
d’Optique's building. The QKD system has been operated with  realistic 
background light, key size in the kbit range and in a configuration where Alice 
and Bob are two entirely remote parties connected via a quantum transmission 
channel in free space  and a classical channel  using internet
 link. 

 In Section 2 we describe the experimental setup used to address single 
color  centers
and the QKD protocol based on polarization encoding on the emitted photons. 
Section 3 deals with the parameters of the QKD experiment. In Section
4, we detail  how the quantum key is extracted  from raw data  using {\sc{Qucrypt}} 
software \cite{Salvail_01}.    Finally, Section
5 is devoted to the discussion of security models for absolute secrecy. We 
will show that 
 in a realistic regime corresponding to   high losses in the quantum 
transmission channel,  our single photon QKD setup 
has a measurable advantage over similar systems using WCP.

\section{Experimental setup}

 \begin{figure}[h]
\includegraphics[width=13cm]{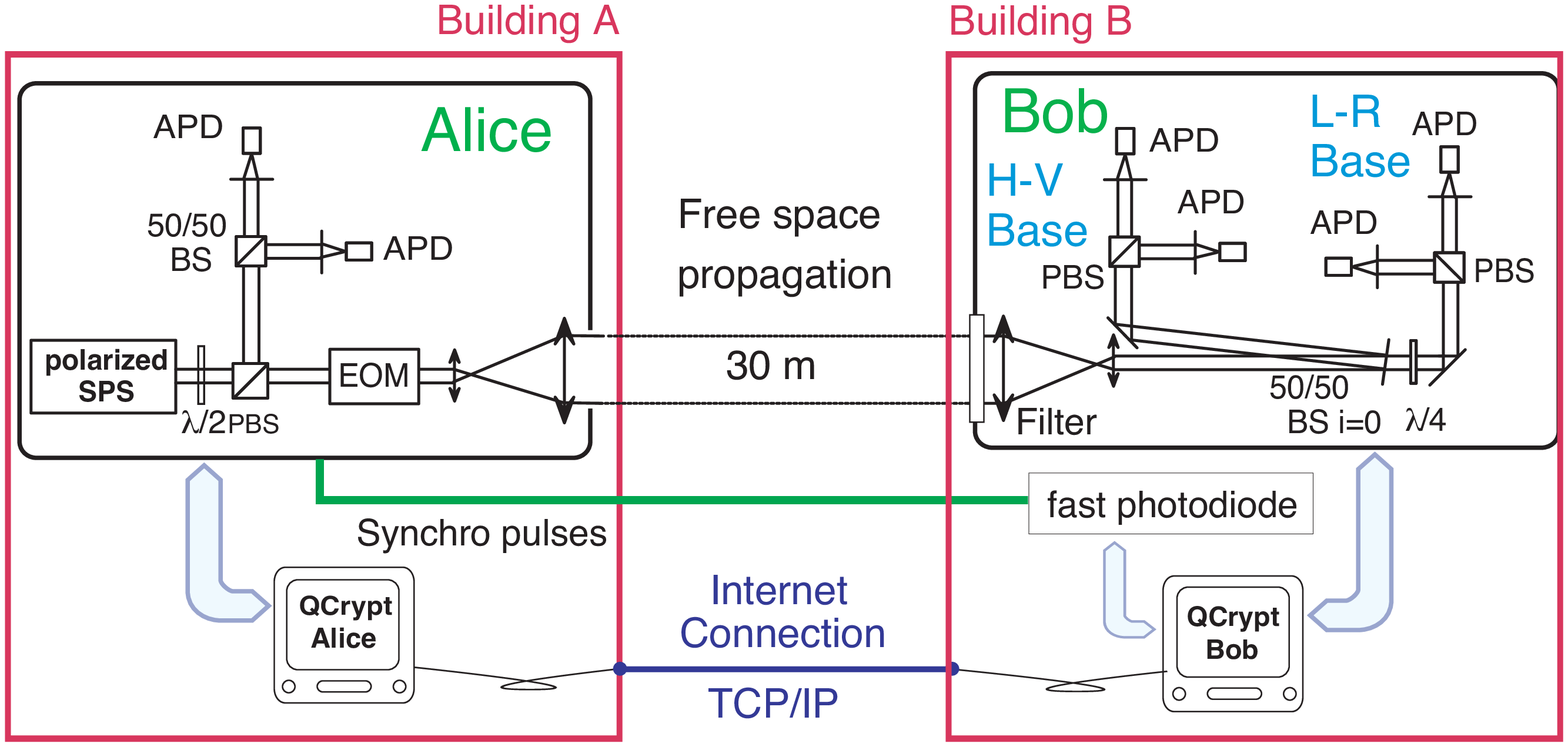}
\caption{{\bf Experimental setup for our quantum key distribution system based on a polarized single photon source. This system corresponds to the implementation of the BB84 protocol. It was operated at night, using a free space quantum channel between Alice and Bob and the Internet as the classical channel. APD: silicon avalanche photodiode. BS: beam splitter. PBS: polarizing beam splitter. EOM: electro optical modulator. $\lambda$/2: achromatic half-wave plate.  $\lambda$/4: achromatic quarter-wave plate}}
\label{expsetup}
\end{figure}

\subsection{Single photon emission}

Lots of effort have been put  in the realization of single photon sources over the 
recent years.  Since first proposals \cite{Yamamoto_heralded,DeMartini_96, 
Rosa_PRA00}, a  great variety of schemes have been worked out, based upon the 
control of fluorescence from  different kind of  emitters, like  molecules 
\cite{Brunel_PRL99,Lounis00,FMT_PRL02}, atoms  \cite{Kuhn_02}, color center  
\cite{Alexios_EPJD}  or   semiconductor structures 
\cite{Michler_Science_00, Santori_PRL_01,
Moreau_APL_01, 
Santori_NAT02,Pelton_PRL02,Izo_APL_03,Vuckovic_APL03,Yuan_02, Alexios_EPJD}. Our   single 
photon source  is based upon the pulsed excitation of a single NV color 
center \cite{Kurtsiefer_00,Rosa00} inside a  diamond nanocrystal 
\cite{Alexios_PRA,Alexios_EPJD}. This type of emitter, which shares many 
similarities with the emission from
 molecules,  has   important practical advantages since  it can be operated 
at room temperature and is    perfectly photostable for both cw and pulsed 
nanosecond excitation\footnote{Note that under femtosecond  pulsed 
excitation,   we observed the photoinduced creation of new color centers 
\cite{Yannick_JLumi_04} in nanocrystal containing initially a single NV 
center. Such behavior   under femtosecond laser illumination place some 
limitations on the use of
sub-picosecond pulses to trigger single photon emission.}.

The nanostructured samples are prepared following a procedure described in 
Ref.\cite{Alexios_PRA}, starting from type Ib synthetic powder (de Beers, 
Netherlands).  
The diamond nanocrystals are size-selected by centrifugation, yielding a 
mean diameter of  about 90 nm.    A polymer solution (polyvinylpyrrolidone, 
1 \% weight in propanol) containing selected diamond nanocrystals is 
deposited by spin-coating on a dielectric mirror, resulting in a 30~nm thick 
polymer layer holding the nanocrystals. The ultra-low fluorescing  
dielectric SiO$_2$/Nb$_2$O$_5$ mirrors (Layertec, Germany)  have been 
optimized to efficiently reflect the emission spectrum of a NV color center, which 
is centered on 690~nm (60~nm FWHM).  Background fluorescence around the 
emission of a single  NV  color center is moreover strongly reduced by photobleaching after a few 
hours of laser illumination, its emission properties remaining unaffected.

Under pulsed excitation with  pulse duration shorter than the excited state 
lifetime (which for the considered samples of NV color centers is 
distributed around 25~ns~\cite{Alexios_PRA}), a single dipole emits single 
photons one by one \cite{DeMartini_96,Rosa_PRA00}. As described in 
Ref.~\cite{Alexios_EPJD}, we use a homebuilt pulsed laser at 532~nm with a 
0.8~ns  pulse duration   to   excite   single NV color center.  The 50~pJ 
energy  per pulse   is high enough  to ensure efficient pumping of the 
emitting center in its excited state. Repetition rate was set to 5.3~MHz so 
that successive fluorescent decays are well separated in time. The green 
excitation light is focused on the nanocrystals by a high numerical aperture  
($NA = 0.95$) metallographic objective. Fluorescence light is collected by 
the same objective. A long-pass filter (low cutting edge at 645~nm) is used 
to block reflected 532~nm pump light.
The stream of collected  photons   is then spatially filtered  by focusing  
into a $100~\mu$m diameter pinhole which ensures the setup confocality.  
Linear polarization of 
the emitted photons is obtained by passing light through a polarizing cube.   
Since the   fluorescence light emitted by a single color center  is  
partially polarized, an  achromatic  half-wave plate is introduced in front 
of the cube. Its rotation allows   to send that linearly polarized fraction 
of the NV fluorescence either towards Bob, or towards two avalanche silicon 
photodiodes (APDs) arranged in a Hanbury Brown and Twiss configuration. This setup  is used to acquire an  histogram of the delay between two 
consecutively detected photons (cf figure \ref{g2_061001}),  from which we infer how far the source 
departs from an ideal SPS.

\subsection{Implementation of the ``BB84'' QKD protocol}

We then  implement the ``BB84'' QKD  protocol, by coding the bits on 
polarization states of the single photons.  We use the horizontal-vertical 
($H-V$) and circular left-circular right
($L-R$) polarization basis.  Each of these polarization states is obtained 
by  applying  a given level of high voltage  on a KDP electro-optical 
modulator  (EOM, Linos LM0202, Germany).  Homemade electronics provides fast 
driving of the high voltage,  being  capable of switching the 300 V halfwave voltage of the EOM within 
30~ns. In our key distribution, the sequence of encoded polarization bits is 
generated with hardware electronics, using two  programmable electronic 
linear shift registers in the Fibonacci configuration. Each register gives a 
pseudo-random  sequence of $2^{20}-1=1048575$ bits, and the ``BB84''  four states 
are coded with two bits,  each of them belonging to one of the two pseudo-random sequences.

As shown on figure \ref{expsetup},  quantum key distribution is realized  
between two parties, Alice and Bob, located in two remote wings of Institut 
d'Optique building (Orsay, France).
Single photons are sent through the windows, from one building to another. To minimize diffraction 
effects,  the beam is enlarged to a diameter of about 2~cm with an afocal 
setup made of two lenses, before sending it through 30.5~m of open air. 
Transmitted photons are   collected on Bob's side by a similar afocal setup 
which reduces its diameter back to the original one.

On Bob's side, a combination of four Si-APDs  was  
used to measure the polarization sent by Alice (see figure \ref{expsetup}). The $ H-V$ or $L-R$  basis 
is passively selected, as the single photons are either transmitted or 
reflected on a 50/50 beam splitter used   at almost $0^\circ$  incidence to avoid any mixing between the four polarization states. In 
the linear polarization  detection basis $H-V$,  states $H$ and $V$ are 
simply  discriminated by a polarizing beamsplitter whose  outputs are sent 
onto two APDs. For the circular $L-R$ basis, an achromatic quater-wave plate  
transforms  the incoming circular polarisations into linear ones, which are finally 
detected with a polarizing beamsplitter and two APDs.  

 The polarization state associated to each detection event on Bob's APDs  is recorded by 
a high speed digital I/O PCI computer card (National Instrument, PCI-6534). 
In order to remove non-synchroneous APD 
dark counts, reading of each detector output is synchronized with the 
excitation pulses. Since the pumping laser is driven by a stable external 
clock, this synchronisation is achieved first by sending a small fraction of 
the excitation laser pulses toward a fast photodiode on Bob's side. The 
photodiode output is reshaped into a 30~ns TTL-like pulse which is 
electronically delayed while the output electric pulses from each APD are reshaped to a constant 60~ns 
duration TTL-like pulse, eliminating any   APD pulse width fluctuation.
The acquisition card reads its states inputs on the 
falling edge of the synchronization pulse. Optimal setting of the 
electronic delay therefore ends up  in a time-gated measurement of the APD 
outputs, within a gate of 60~ns width.

The sequence of   time-gated polarization state measurements   constitutes 
Bob's raw key. It can be considered as the output of the  ``quantum 
communication phase" which lasts a period of 0.2~s.
The remaining steps of the ``BB84'' QKD  protocol are purely classical 
ones. They consist in taking advantage of the quantum correlations between 
Alice's  information and Bob's raw key   in order to distill   secrecy 
between these two parties. All theses steps, detailed in Part 
\ref{distillation}, are realized over the internet using TCP/IP protocol, by 
the open source {\sc{Qucrypt}} software written by L.~Salvail (Aahrus 
University, Denmark)  \cite{Salvail_01}.

\section{Parameters of the QKD experiment }
\label{parameters}

The principal goal of our experiment was to bring together a realistic setup 
in order to test   practical feasibility of   single-photon open air QKD. 
Experimental sessions were done   from the end of August 2003 to the middle 
of September 2003. The system was operated at night so as to keep the 
influence of background light (in our case,  moon and public lightning) at a 
relatively low level. 
 Our room temperature SPS proved its convenience 
and reliability in these experimental conditions.  Note that for  
consistency reasons, all the data 
analyzed  in the article  were  obtained from the emission of a  given 
single NV color center, chosen for  its  strong emission rate. Keeping always this same center allows to consistently investigate the effect of  high attenuation on the quantum transmission 
channel.  

\subsection{Emission efficiency  of the SPS and  assessment  of its subpoissonian statistics}

Preliminary   characterization of the SPS  quality,   performed on Alice's 
side, consists in  measurements  of   the emission rate and   the reduction 
in   probability of multiphotonic emissions, compared with an equivalent WCP of same mean 
number of photons per pulse.

For a 0.2 s sequence of transmission and a pulsed excitation of $5.3 \; {\rm MHz}$  a total  of $8.8 \times 10^{4}$ photons is recorded on Alice's side. 
 By correcting from the APD efficiency $\eta_{\rm APD}=0.6$, we   can thus infer  that    
the overall emission efficiency of the polarized SPS is of about $\approx 2.8 \%$.
After polarization encoding in the EOM of  transmission $T_{\rm EOM}=0.90$  
and transmission $T_{\rm optics}=0.94$ through the optics of the telescope,
  the mean number of polarized single photon per pulse  sent on the quantum 
channel is $\mu = 0.0235$.

\begin{figure}[h]
 \includegraphics[width=12cm]{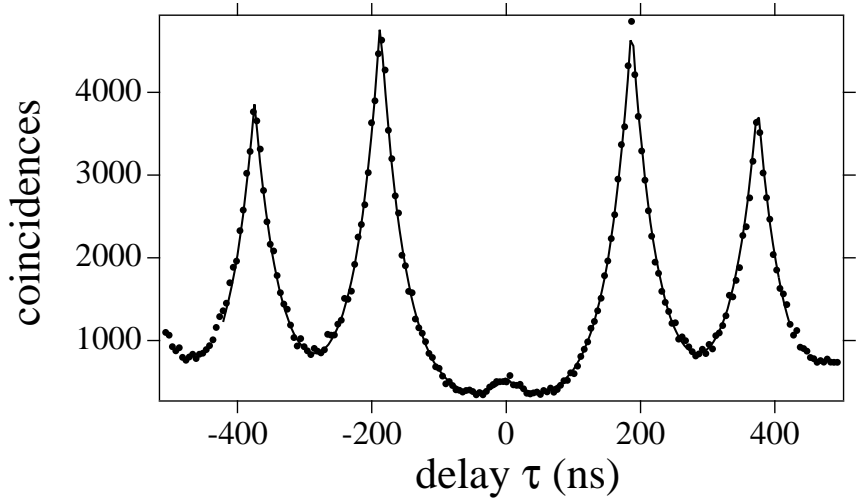}
\caption{{\bf Histogram of time intervals between consecutive photon detection events in Alice's correlation setup. Integration time is 175 s. Lines are exponential fits for each peak, taking into account background level. Radiative lifetime given by the fit is 35 ns and the repetition period is of 188 ns. The strong reduction of coincidences at zero delay gives evidence for single photon emission by the excited color center.
}}
\label{g2_061001}
\end{figure}

Direct evidence for the reduction of multiphotonic emission probability comes from the acquisition 
of the delays with the Hanbury Brown and Twiss setup on Alice's side (figure 
\ref{g2_061001}).
The photon  statistics of the SPS can be  quantified more precisely from  Bob's 
measurements  
which give the probability distribution of the number of photocounts within 
the 60 ns timeslots used for time-gated detection. To perform such 
evaluation,  we have gathered the data corresponding to more than $40 \times 
10^{6}$ pulses registered  by Bob's
acquisition card. For a given  detection timeslot probabilities for 
detecting one and two photons are respectively $P_{d}(1) = 7.6 \times 
10^{-3}$ and $P_{d}(2) = 2.7 \times 10^{-6}$.
From these numbers, we can infer the amount of reduction of   multiphotonic 
emission probability with respect to the photon statistics of an equivalent WCP  \cite{Rosa_PRA00}. 
Note that one needs to take into account the fact that each avalanche photodiode cannot detect more than one photon per timeslot, due to their detection deadtime.
From the configuration of the APDs detection scheme on Bob's side, the
probability $P_d(2)$ to detect two photons is only $5/8$ of the probability
for Bob to receive two photons, the probability that two photons arrive on
the same APD being $3/8$.

Reduction factor $\cal R$ of the multiphotonic probability is therefore  
\begin{equation}
{\cal R} = \frac{5}{8} \times  \frac{P_{d}(1)^{2}/2}{   P_{d}(2)}= 6.7  \; . 
\end{equation}
That result agrees well with  the sub-poissonnian reduction factor of $6.1$
that can be inferred from the normalized area of figure~\ref{g2_061001}, 
taking into account the 60 ns integration time and the lifetime of the 
emitter \cite{Alexios_EPJD}. 
For security analysis and numerical simulations, a  value of ${\cal R} 
= 6.7$ for the sub-poissonian reduction factor will be taken since it 
corresponds to a  direct  outcome of the photocounts record.

As it will be   discussed in  more details in the section concerning security 
models,   information leakage towards   potential eavesdropper is directly 
linked to $S^{({\rm m})}$ which is the probability per excitation pulse that a multiphotonic pulse leaves on Alice's 
side. For the equivalent WCP, that parameter is:
\begin{equation}
S^{({\rm m})}_{{\rm WCP}} =1-(1+\mu) {\rm e}^{-\mu}= 2.7 \times 10^{-4}
\end{equation}
whereas for  the  SPS, that parameter can be evaluated as     
\begin{equation}
S^{({\rm m})}_{{\rm SPS}} =\frac{1}{6.7} \,  \left\lbrack 1-(1+\mu) {\rm 
e}^{-\mu} \right\rbrack
= 4.1 \times 10^{-5}   \; .  
\end{equation}

\subsection{Parameters of Bob's detection apparatus } 
Probabilities for recording a photocount on one of Bob's detectors within 
a given timeslot is $p_{\rm exp} \simeq  7.6 \times 10^{-3}$. Making the 
reasonable assumption that absorption in the 30 m open-air transmission beam is 
negligeable  and taking into account the $\mu=0.0235 $ value, one can infer an estimate 
of the efficiency of Bob's detection apparatus  as $\eta_{\rm Bob} \simeq 0.3$

Detector dark counts and fake photocounts due to stray light
are  responsible on Bob's side for errors in the key exchange process. As it  
will be discussed in more details below, these errors  contribute to the 
practical limit of secure transmission distance.  Particular care was taken to protect Bob's APDs from 
stray light, using shielding and spectral filtering. Nevertheless, due to  the broad 
emission spectrum of NV color centers, benefit of spectral filtering remains limited and the 
experiment could not be runned under usual day light conditions.  At night, 
the measured dark count rates on Bob's APDs (in experimental conditions 
after stopping the SPS beam),  $(d_{H}, d_{V}, d_{L}, d_{R})$ are (60,70, 
350, 150) s$^{-1}$. 
Considering the ratio of the 60 ns detection timeslots compared to the 35 ns radiative lifetime of the NV color center,  $82 \%$ of the SPS photons are falling within the detection gate while only $32 \%$ of the dark counts are introduced in the key exchange process. The probability  of  a dark count record within  a given  detection timeslot is thus $ p_{\rm dark}=3.8 \times 10^{-5} \, \,  {\rm{s}}^{-1}$.

\subsection{Evaluation of   quantum bit error rate}
Quantum bit error rate (QBER) is computed by comparing  Alice and Bob's data corresponding to same polarization basis.  The errors are due to two experimental imperfections of the system. First, non ideal polarization encoding and detection can result in optically induced errors, which number is proportional to detected signal level. 
Second, dark counts of the APDs induce errors within the transmission sequence, 
which average number is independent of the mean number of photon per pulse. 

Following the analysis of Ref.~\cite{Lutkenhaus_00} and accounting for the 
specificities
of our detection setup on Bob's side, QBER  $e$ is given by 
 \begin{equation}
e= \alpha  \frac{p_{\rm signal}}{p_{\rm exp}} + \frac{ p_{\rm dark}}{ p_{\rm exp}}
\label{eq_error_rate}
\end{equation}
where $\alpha$ an adjustable parameter and
$p_{\rm signal}$ is the probability to detect a signal photon independently of dark count.  $p_{\rm signal}$ is an estimated value, based on a calculation taking into account $\mu$ as well as the value of the attenuation on the quantum channel.
  Measurements of QBER $e$ for different 
attenuation values of the quantum transmission channel  (i.e. variations of $ 
p_{\rm signal}$ and 
$p_{\rm exp}$) are given in Table~\ref{table_attenuation_erreur}.  Measured 
values 
of $e \times p_{\rm exp}$  correlate  well with $p_{\rm signal}$ values, 
accordingly to Eq.~(\ref{eq_error_rate}). Linear fit gives $\alpha = 1.3 
\times 10^{-2}$ and
$p_{\rm dark} = (35 \pm 6) \times 10^{-6}$, a result compatible with 
previous direct estimate. These values will be used in all following 
numerical simulations.

\begin{table}
\begin{tabular}{c c c c c}
\hline \hline
Added attenuation & Average size of raw data (bits) & $p_{\rm exp} $  &  QBER \\
1   	&		8000			&  $7.6 \times 10^{-3}$    &  1.65 $\%$  \\
0.498  	&		4250		& $4.0 \times 10^{-3}$   &   2.2 $\%$  \\
0.25   &		2100			& $2.0 \times 10^{-3}$   &   3.2 $\%$ \\
0.128   &		1025			& $9.8 \times 10^{-4}$   &   4.15 $\%$ \\
0.057   &		395			& $3.8 \times 10^{-4}$   &     9.4 $\%$ \\
\hline \hline
\end{tabular}
\caption{Measured experimental parameters as a function of the added attenuation on the quantum channel.
In order to limit statistical fluctuations, values of the QBER $e$ and of $p_{\rm exp} $ have been computed on samples of at least 3000 bits, obtained by concatenation of several raw data samples.  } 
\label{table_attenuation_erreur}
\end{table}

\section{Experimental implementation of ``BB84'' QKD protocol}

\subsection{Raw key exchange  and sifted data}

During a key transmission sequence lasting 0.2 s, Bob detects 
approximatively  a fraction $\eta_{\rm{Bob}} \times \mu$ of the 1048575 bits 
initially  encoded by Alice. 
Without any added attenuation on the quantum transmission channel, Bob 
detects on average $8000$ bits, which constitute the initial raw data exchanged
through the physical quantum channel.  Starting from this shared 
information, Alice and Bob then extract a key by 
exchanging classical information for basis reconciliation. Bob reveals the 
index of the pulses for which a photocount has been recorded and publicly announces to which 
polarization basis ($H-V$ or $L-R$) it belongs. Events corresponding to more than 
one photodection on Bob's APDs are discarded  since they are
ambiguous.  Note that 
one should nevertheless impose an upper bound on the acceptable number for 
such events, so that discarding them does not introduce a backdoor for any 
eavesdropper. 
Considering the low number of multiple photodetection events in our 
experiment, 
such filtering does not introduce any practical  limitation in the key distillation process.
Alice then 
reveals which   bits 
correspond to identical polarization basis and should be retained. This 
process
ends up in sifted data, which number  $N_{\rm sifted} \approx 4000$ is on 
average half the number of Bob's recorded data.    

\subsection{Key distillation from sifted data }
\label{distillation}

Sifted data  shared by Alice and Bob have imperfect correlations since they 
are affected by errors. They are moreover not perfectly secure since an 
eavesdropper may have gained some information on exchanged bits during  
the quantum transmission sequence. 

Starting from those data, complete secrecy is then obtained  
by error correction followed by privacy amplification \cite{Bennett_95}.
That two-steps procedure, which allows one to distill a secret key for the sifted data, is achieved throughout the IP network using the public domain software {\sc{Qucrypt}} \cite{Salvail_01}.
 
{\sc{Qucrypt}} uses the algorithm {\sc{Cascade}} for error correction \cite{Brassard_93}. 
It implements an iterative dichotomic splitting of Alice and Bob 
sifted data into blocks and compares their parity in order to spot and 
correct the errors. This algorithm is optimized to correct all the errors  
while revealing a minimum number of bits. For a QBER $e$, the Shannon information 
\begin{equation}
f(e) = -\log_{2}e - (1-e) \log_{2}(1-e)
\end{equation}
gives a lower bound on the amount of information that needs to be exchanged on the public channel to correct one error.

A random subset  of  $1 \%$ of the data used by {\sc Qucrypt}  is taken to evaluate the QBER $e$. 
With such length of tested data the number of secure bits extracted from the  sifted data 
fluctuates by less than 5\%   from  one run of {\sc Qucrypt} to another.
Moreover, for our data samples of a  few thousands of bits,  {\sc{Cascade}} 
corrects errors with a good efficiency. We indeed checked that the 
information disclosed to correct one error is only $10 \%$ greater than the 
limit imposed by the  Shannon bound.

The total amount of information an eavesdropper may have gained on the sifted data is  a crucial parameter for the final privacy amplification step.
It is the sum of two contributions: the information classically disclosed during error correction added to the information that Eve may have gained  during the quantum transmission. This later part has to be evaluated accordingly to security requirement and model. 

By setting an upper bound on the QBER in data processing by {\sc Qucrypt}  we ensure that all our QKD sessions are secure against a first class of attack. We set this bound to 12.5\%, which corresponds to the minimum probability for Eve to introduce an error by performing measurements on a single pulse without knowing Bob's measurement basis \cite{Polzik_JMO01}. 
However more efficient attacks can be used. We therefore assessed the security of our data in reference to the approach of N. L\"{u}tkenhaus, who developed a theoretical framework for the secure experimental QKD implementations of the ``BB84''  protocol \cite{Lutkenhaus_00}. It has the nice feature of giving a positive security proof for realistic experimental systems under the so-called individual attacks. The calculations are based on the assumption that Eve's optimal strategy is to  perform a Photon Number Splitting (PNS) attack on multiphotonic pulses, allowing her to finally get all the information carried on those pulses. 
 
 Although such strategy might not always be the optimal one \cite{Curty}, it becomes the most efficient eavesdropping scheme on the ``BB84'' protocol for strong transmission losses on the quantum channel. Note that alternative protocols to the``BB84'' protocol, robust against the PNS attack, have been recently proposed \cite{Acin}  and might constitute an efficient way to increase the span of experimental QKD systems relying on WCPs. Under high attenuation, such schemes allow to work with higher $\mu$ (mean number of photons per pulse) since information carried on two-photon pulses is less vulnerable to eavesdropping.  It would be interesting to compare the performance of SPSs  with respect to WCPs in this case, although such analysis is beyond the scope of the present article.

\section{ Performance of the QKD setup and resistance to losses}

Secure key distribution performance of the QKD system is characterized by 
the mean amount  of secure information exchanged on each sent pulse.
Experimental measurements of that parameter have been performed for different level of losses in the quantum channel. The results are compared to numerical simulations based on the analytical derivation of the 
number of secure bits per pulse $G$ after privacy amplification and error 
correction evaluated from the analysis of Ref.  \cite{Lutkenhaus_00} and given by 

\begin{eqnarray}
\kern -2cm
\lefteqn{G =  \frac{1}{2}
 \,p_{\rm exp}   \Bigg\{  \frac{p_{\rm exp}-S^{(\rm m)}} {p_{\rm exp}} \left( 1-
log \left[  1 + 4 e   \frac{p_{\rm exp}}{p_{\rm exp}-S^{(\rm m)}}  - 4 \left(e
\frac{p_{\rm exp}}{p_{\rm exp}-S^{(\rm m)}}\right)  ^{2}\right]  \right) } \nonumber \\
   +   1.1 \,  \left[ \log_{2}e + (1-e) \log_{2}(1-e) \right]  \Bigg\}
\label{Gformula}
\end{eqnarray}

\begin{figure}[h]
 \includegraphics[width=11cm]{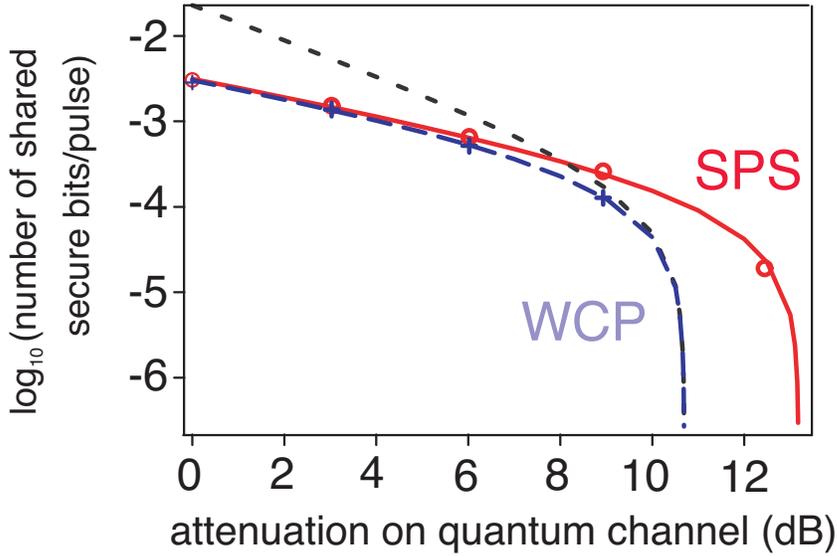}
\caption{{\bf Simulation and experimental data for the number of exchanged secure bits 
per time slot, versus attenuation in the quantum channel.
Solid-line red curve and wide-dashed blue curve correspond respectively to 
numerical simulations of equation (\ref{Gformula}) for SPS and WCP, using experimental parameters given in 
section (\ref{parameters}). The narrow-dashed curve is obtained by 
optimizing $G$ with respect to $\mu$ in equation (\ref{Gformula}). It 
corresponds to the limit of WCP performance under our experimental 
conditions and security model. 
}}
\label{simulatt}
\end{figure}

Theoretical curves giving $G$ versus attenuation on the quantum transmission channel are 
displayed on figure \ref{simulatt}, together with experimental points corresponding either to our SPS or to an equivalent WCP. Measured experimental rates correspond to data samples 
large enough to ensure a statistical accuracy better than $5$ $\%$. 
They are in good agreement with theoretical curves showing that 
experimental parameters have been correctly assessed and that data samples 
are large enough for efficient error correction

In the absence of attenuation, an average of $3200$ secure bits can be exchanged within the 0.2 s transmission sequence. It corresponds to a $16$ kbits/s rate,  twice larger than the one of the first 
experimental realization \cite{SPS_alex}. 
As it can be seen on figure \ref{simulatt}, reduction of the proportion 
of multiphotonic pulses gives a significant advantage of our system over 
WCP, in the strong attenuation regime. Since our setup is affected by 
relatively high level of dark counts 
\footnote{
There are several reasons for that. The main one is inherent to the long 
emission liftetime of our SPS, forcing us to use long (here 60 ns) detection 
window. There are two other reasons that coud be subject to improvement : we are 
using a passive determination of the detection basis on Bob side, which 
increases dark counts by a factor of two, and two of our Si 
APDs have dark count rate higher than the common value of $70$ Hz.}
 and since we have adopted a restrictive security model, our system cannot work under attenuation stronger than 13 dB. It however allows us to directly check for the influence of the photon statistics on the experimental QKD system.

A first comparison consists in keeping a constant value of $\mu=0.0235$ and calculating 
the effect of either sub-poissonian or poissonian statistics on the size of the final key. This directly relates  to the comparison of the ``SPS'' and ``WCP' curves on figure \ref{simulatt}.
When the system is operated with WCP, one can try to optimize $G$ over $\mu$ for different attenuation values. 
However, even with this strategy (cf figure 
\ref{simulatt}) it clearly appears that our SPS overcomes WCP
operated in same experimental conditions, as soon as attenuation 
reaches 9 dB. In all cases the maximum distance at which secure key distribution 
can be guaranteed is increased by more than 2 dB.

\section{Conclusion}

In this paper we have demonstrated a free space QKD setup using the ``BB84''
protocol. The system is based on a stable, simple, and reliable pulsed single photon source
(SPS). The open air experimental conditions in which it was operated  are reasonably close  to those 
for practical application. They might be extended to kilometric distances using previously established techniques \cite{Kurtsiefer_03}. Advantages of SPS over equivalent weak coherent pulses (WCP)   have been experimentally assessed for increasing propagation losses. The results
demonstrate quantitatively that QKD with SPS outperforms QKD with WCP, when transmission losses exceed 10 dB.

There clearly remains much room for improvement. For instance, SPS
sources using quantum dots \cite{Santori_NAT02, Pelton_PRL02,Izo_APL_03,Vuckovic_APL03} are able to emit much shorter pulses with much narrower bandwidths than   diamond NV color centers. Those properties are indeeed very
favorable for efficient QKD but presently
require  a cryogenic (liquid He) environment. This constraint   makes quantum-dot-based QKD   much
less suitable for outdoor applications than our SPS. Avenues might be found either by developing semiconductor quantum dots operating at higher temperature
(e.g. with II-VI semiconductors), or by finding other color centers with
improved performances. New improvement can also be foreseen on the protocol
side \cite{Acin}, where both SPS and non-SPS sources deserve to be examined.

Presently, neither  color-center- nor quantum-dot-based SPS can operate at the
telecom wavelength range around 1550 nm. Their main application given their
emission wavelength   is free-space QKD, especially  QKD from
satellite \cite{Rarity_02}. Compactness and reliability then become major issues.
Development of  nanofabrication techniques should allow the realization of
compact sources based on diamond nanocrystals. In any case, QKD systems have in recent years
overcome many difficulties initially considered insurmountable. It is promising that such  progress will continue in the near future.

\section*{Acknowledgements}
We thank Thierry Gacoin for realizing the NV centers samples and Louis Salvail and Martial Tarizzo for help with the {\sc QUCRYPT} software. This work was supported by the European Commission (IST/FET program), by France Telecom R$\&$D and by the "ACI Jeunes Chercheurs"  (Ministère de la Recherche et des Nouvelles Technologies).

\end{document}